\pdfoutput=1

\documentclass[iop]{emulateapj}

\slugcomment{}

\shorttitle{Hub-filament system in IRAS 05480+2545: young stellar cluster and 6.7 GHz methanol maser}
\shortauthors{L.~K. Dewangan}
\begin{document}

\title{Hub-filament system in IRAS 05480+2545: young stellar cluster and 6.7 GHz methanol maser}
\author{L.~K. Dewangan\altaffilmark{1}, D.~K. Ojha\altaffilmark{2}, and T. Baug\altaffilmark{2}}
\email{lokeshd@prl.res.in}
\altaffiltext{1}{Physical Research Laboratory, Navrangpura, Ahmedabad - 380 009, India.}
\altaffiltext{2}{Department of Astronomy and Astrophysics, Tata Institute of Fundamental Research, Homi Bhabha Road, Mumbai 400 005, India.}
\begin{abstract}
To probe the star formation (SF) process, we present a multi-wavelength study of IRAS 05480+2545 (hereafter I05480+2545). 
Analysis of {\it Herschel} data reveals a massive clump (M$_{clump}$ $\sim$1875 M$_{\odot}$; peak N(H$_{2}$) $\sim$4.8 
$\times$ 10$^{22}$ cm$^{-2}$; A$_{V}$ $\sim$51 mag) containing the 6.7 GHz methanol maser and I05480+2545, which is also 
depicted in a temperature range of 18--26 K. Several noticeable parsec-scale filaments are detected in the {\it Herschel} 250 $\mu$m image and 
seem to be radially directed to the massive clump. It resembles more of a ``hub-filament" system. 
Deeply embedded young stellar objects (YSOs) have been identified using the 1--5 $\mu$m photometric data, 
and a significant fraction of YSOs and their clustering are spatially found toward the massive clump, revealing the intense SF activities. 
An infrared counterpart (IRc) of the maser is investigated in the {\it Spitzer} 3.6--4.5 $\mu$m images. 
The IRc does not appear point-like source and is most likely associated with the molecular outflow. 
Based on the 1.4 GHz and H$\alpha$ continuum images, the ionized emission is absent toward the IRc, indicating that the massive clump harbors an early phase of massive protostar before the onset of 
an ultracompact H\,{\sc ii} region. Together, the I05480+2545 is embedded in a very similar ``hub-filament" system to those seen in Rosette Molecular Cloud. 
The outcome of the present work indicates the role of filaments in the formation of the massive star-forming clump and cluster of YSOs, which might help channel material to the central hub configuration and the clump/core.
\end{abstract}
\section{Introduction}
\label{sec:intro}
The formation of massive stars ($\gtrsim$ 8 M$_{\odot}$) and young stellar clusters is still not well understood. 
In star-forming regions, the filamentary structures often harbor dense massive star-forming clumps and young stellar clusters 
\citep[e.g.][and references therein]{andre10,andre16,schneider12,ragan14,kainulainen16,contreras16,li16}. 
However, the role of filaments in the star formation process (including massive stars) is still a matter of debate 
\citep[e.g.][]{myers09,schneider12,nakamura14,tan14}. 
The filamentary structures may contain H\,{\sc ii} regions and Class~II 6.7-GHz methanol masers \citep[e.g.][]{schneider12,dewangan16b}.
It has been suggested that the 6.7 GHz methanol maser emission (MME) is a reliable tracer of massive young stellar objects (MYSOs) 
\citep[e.g.][]{walsh98,minier01,urquhart13}. 
Therefore, such sites can also be useful to search for massive stars at their early formation
stage prior to an ultracompact (UC) H\,{\sc ii} phase, which is observationally very challenging task \citep[e.g.][]{tan14}.

Located at a distance of 2.1 kpc \citep{henning92,kawamura98,klein05,wu10,wu11}, IRAS 05480+2545 (hereafter I05480+2545) is a very poorly studied star-forming region. 
The previously known H\,{\sc ii} region BFS 48 is located at a projected linear separation of about 55$\arcsec$ (or $\sim$0.55 pc at a distance of 2.1 kpc) 
from I05480+2545. Another IRAS source, IRAS 05480+2544 is linked with the H\,{\sc ii} region BFS 48 \citep{blitz82}. 
Using the NANTEN $^{13}$CO (1-0) emission (beam size $\sim$2$'$.7), \citet{kawamura98} investigated the molecular clouds 
in Gemini and Auriga including the I05480+2545 and BFS 48 sites \citep[see BFS 48 region around $l$ = 183$^{\circ}$.33; 
$b$ = $-$0$^{\circ}$.53 in Figure 9j in][]{kawamura98}. They designated this molecular cloud as ``183.3$-$00.5" 
(V$_{lsr}$ $\sim$$-$9.1 km s$^{-1}$) 
and estimated the mass (radius) of the cloud to be $\sim$3000 M$_{\odot}$ (6 pc) \citep[see source ID \#79 in Table 1 in][]{kawamura98}. 
\citet{klein05} observed millimeter continuum emission at 1.27 mm and an MSX source toward I05480+2545 \citep[see Figure~1 in][]{klein05}. 
A 6.7~GHz MME was reported toward I05480+2545 \citep[designation: 183.35-0.58; peak velocity $\sim$$-$4.89 km s$^{-1}$;][]{szymczak12}. However, the investigation of MYSO(s) toward the 6.7~GHz MME is yet to be carried out. 
There were also other masers (such as, water maser at 22.2 GHz \citep{macleod98,sunada07}, hydroxyl maser at 1665 MHz \citep{slysh97}, and methanol maser at 12.2 GHz \citep{macleod98}) reported in the I05480+2545 site. 
\citet{wu10} presented the NH$_{3}$ and $^{12}$CO (1-0) line observations toward the low- and high-luminosity Class~II 6.7 GHz methanol masers including the I05480+2545 site and reported the detection of a NH$_{3}$ core in I05480+2545 and its physical parameters (such as, density, column density, temperature, core size, and mass) \citep[see Table~7 in][]{wu10}. Using the NH$_{3}$ line data, they also measured the velocity of the dense molecular gas to be $\sim$ $-$9.64 km s$^{-1}$, which is consistent with velocity obtained by $^{13}$CO gas \citep[e.g.][]{kawamura98}. 
Furthermore, \citet{wu11} carried out multi-line observations, including transitions of $^{13}$CO(1-0), C$^{18}$O(1-0), HCO$^{+}$(1-0) and 
N2H$^{+}$(1-0), towards nine low-luminosity 6.7-GHz methanol masers including the I05480+2545 site and reported 
the physical parameters of the core observed in this site \citep[see Table~7 in][]{wu11}.
However, to our knowledge, the study of physical conditions in the I05480+2545 site at large scale is still limited. 
The identification of filaments and the investigation of young stellar clusters are unknown in the molecular cloud linked with I05480+2545 (hereafter MCI05). 
In this paper, to understand the ongoing physical processes in the MCI05, 
we have carried out a detailed multi-wavelength study of observations from optical H$\alpha$, near-infrared (NIR) to radio wavelengths.  

In Section~\ref{sec:obser}, we provide the description of the adopted data-sets in this paper. 
In Section~\ref{sec:data}, we give the results concerning the physical environment and point-like sources.  
The possible star formation processes are presented in Section~\ref{sec:disc} . 
Finally, the major outcomes are summarized in Section~\ref{sec:conc}.
\section{Data and analysis}
\label{sec:obser}
The target field around I05480+2545 considered in this work is chosen based on the the extent of the 
molecular cloud ``183.3$-$00.5" as traced in the $^{13}$CO map \citep[see Figure 9j in][]{kawamura98}. 
The field has a size of $\sim$0$\degr$.55 
$\times$ 0$\degr$.55 ($\sim$20.1 pc $\times$ 20.1 pc at a distance of 2.1 kpc) 
and is centered at $l$ = 183$\degr$.247, $b$ = $-$0$\degr$.605. 
In the following, a brief description of each of the adopted data sets is presented.
\subsection{Narrow-band H$\alpha$ image}
Narrow-band H$\alpha$ image at 0.6563 $\mu$m was extracted from the Isaac Newton Telescope Photometric H$\alpha$ Survey of the Northern Galactic Plane \citep[IPHAS;][]{drew05}. The survey was conducted using the Wide-Field Camera (WFC) at the 2.5-m Isaac Newton Telescope, located at La Palma. 
The WFC contains four 4k $\times$ 2k CCDs, in an L-shape configuration. The pixel
scale is $0\farcs33$ and the instantaneous field of view is about 0.3 square degrees. 
More details of the IPHAS can be obtained in \citet{drew05}. 
\subsection{NIR (1--5 $\mu$m) Data}
We utilized the deep NIR photometric {\it HK} magnitudes of point sources extracted from the UKIDSS Galactic Plane Survey \citep[GPS;][]{lawrence07} tenth archival data release (UKIDSSDR10plus) and Two Micron All Sky Survey \citep[2MASS;][]{skrutskie06}. The GPS K band image was also obtained. 
The UKIDSS observations (resolution $\sim$$0\farcs8$) were performed with the Wide Field Camera (WFCAM) mounted on the United Kingdom Infra-Red Telescope. 
2MASS photometric data were used to calibrate the final fluxes. 
In this work, we obtained only a reliable NIR photometric catalog. 
More information about the selection methods of the GPS photometry can be found in \citet{dewangan15}.
Our resultant GPS catalog contains point sources fainter than H = 12.3 and K = 11.7 mag to avoid saturation. 
In our final catalog, the magnitudes of the saturated bright sources were replaced by the 2MASS photometric magnitudes. 
To depict reliable 2MASS photometric data, we retrieved only those sources having photometric magnitude error of 0.1 or less in each band.

Narrow-band H$_{2}$ ($\nu$ = $1-0$ S(1) at $\lambda$ = 2.122 $\mu$m) imaging observations of I05480+2545 were made on 16 April 2017 
using the Near-Infrared Camera and Spectrograph (NICS: \citet{anandarao08}) mounted on the 1.2m telescope of the Mount Abu InfraRed Observatory (MIRO).
NICS has a 1024 $\times$ 1024 HgCdTe array Wide-area Infrared Imager-I (HAWAII-1; Teledyne, USA).
With a plate scale of $\sim$0\farcs5/pixel, it covers a field of view of $\sim$8$'\times$8$'$ on the sky. 
Apart from the observations of the programmed object, we also observed near-by sky with same exposure time for sky subtraction. 
These images were reduced using the standard analysis procedure available in IRAF and STARLINK packages, like dark and sky subtraction. 
Seventeen images were co-added to get final programmed object image, where an exposure of each single frame was 90 seconds. 
The total integration time for the co-added images is 1530 seconds.
During the observations, an average seeing was estimated to be about 2$\arcsec$. 
The astrometry of final processed image was calibrated using the 2MASS K-band point sources.

Warm-{\it Spitzer} IRAC 3.6 and 4.5 $\mu$m photometric images (resolution $\sim$2$\arcsec$) 
and magnitudes of point sources were downloaded from the Glimpse360\footnote[1]{http://www.astro.wisc.edu/sirtf/glimpse360/} \citep{whitney11} survey.
We extracted the photometric magnitudes from the Glimpse360 highly reliable catalog. 
To select further reliable Glimpse360 photometric data, we obtained only those sources having photometric magnitude error of 0.2 or less in each band.
\subsection{Mid-infrared (12--22 $\mu$m) Data}
We downloaded mid-infrared (MIR) images at 12 $\mu$m (spatial resolution $\sim$6$\arcsec$) and 22 $\mu$m (spatial resolution $\sim$1$2\arcsec$) 
from the publicly available archival WISE\footnote[2]{Wide Field Infrared Survey Explorer, which is a joint project of the
University of California and the JPL, Caltech, funded by the NASA.} \citep{wright10} database.
\subsection{Far-infrared and Sub-millimeter Data}
\label{subsec:her}
Far-infrared (FIR) and sub-millimeter (sub-mm) continuum images were retrieved from the {\it Herschel} Space Observatory data archives.
The processed level2$_{-}$5 images at 70 $\mu$m, 160 $\mu$m, 250 $\mu$m, 350 $\mu$m, and 500 $\mu$m were 
obtained using the {\it Herschel} Interactive Processing Environment \citep[HIPE,][]{ott10}. 
The beam sizes of the {\it Herschel} images at 70 $\mu$m, 160 $\mu$m, 250 $\mu$m, 350 $\mu$m, 
and 500 $\mu$m are 5$\farcs$8, 12$\arcsec$, 18$\arcsec$, 25$\arcsec$, and 37$\arcsec$, respectively \citep{poglitsch10,griffin10}. 
\subsection{Dust continuum 1.1 mm data}
We also used the dust continuum 1.1 mm map \citep{aguirre11} 
from the Bolocam Galactic Plane Survey (BGPS). 
The effective full width at half maximum (FWHM) of the 1.1 mm map is $\sim$33$\arcsec$. 
\subsection{Radio continuum data}
The radio continuum map at 1.4 GHz (21 cm; beam size  $\sim$45$\arcsec$) was retrieved from the NRAO VLA Sky Survey \citep[NVSS;][]{condon98} archive.
The radio continuum emission is useful to trace the ionized emission.
\subsection{H\,{\sc i} line data}
We also used H\,{\sc i} line data at 21 cm from the Canadian Galactic Plane Survey \citep[CGPS;][]{taylor03}.
The velocity resolution of H\,{\sc i} line data is 1.32~km\,s$^{-1}$, sampled every
0.82~km\,s$^{-1}$. The data have a spatial resolution of 1$\arcmin$ $\times$ 1$\arcmin$ csc$\delta$.
The line data have a brightness-temperature sensitivity of $\Delta$T$_{B}$ = 3.5 sin$\delta$ K.
More details of the CGPS observing and data processing strategy can be found in \citet{taylor03}.
\section{Results}
\label{sec:data}
\subsection{Physical environment of I05480+2545}
\label{sec:env}
\subsubsection{Large-scale view of I05480+2545}
\label{sec:envl}
\begin{figure*}
\epsscale{0.535}
\plotone{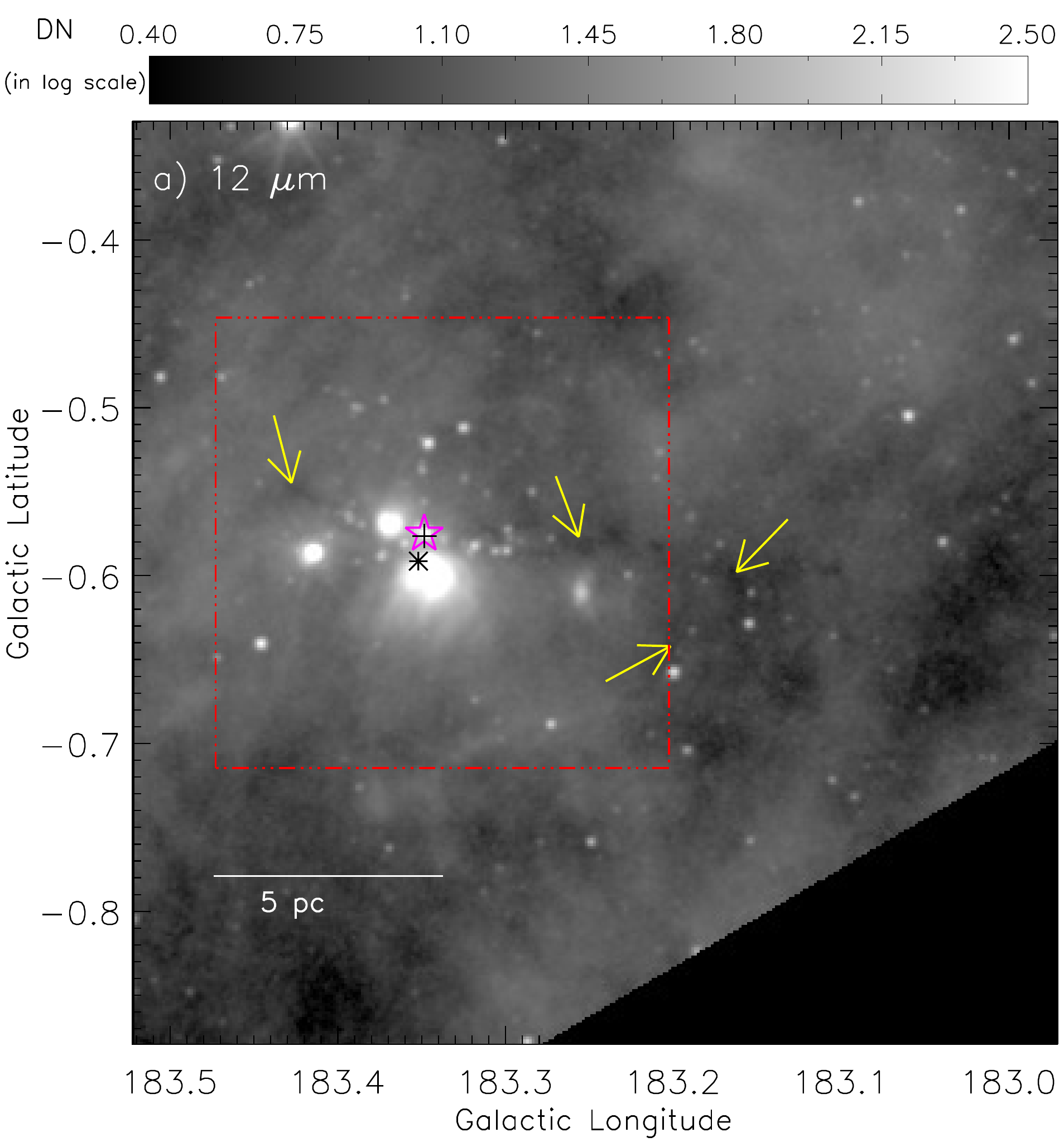}
\epsscale{0.535}
\plotone{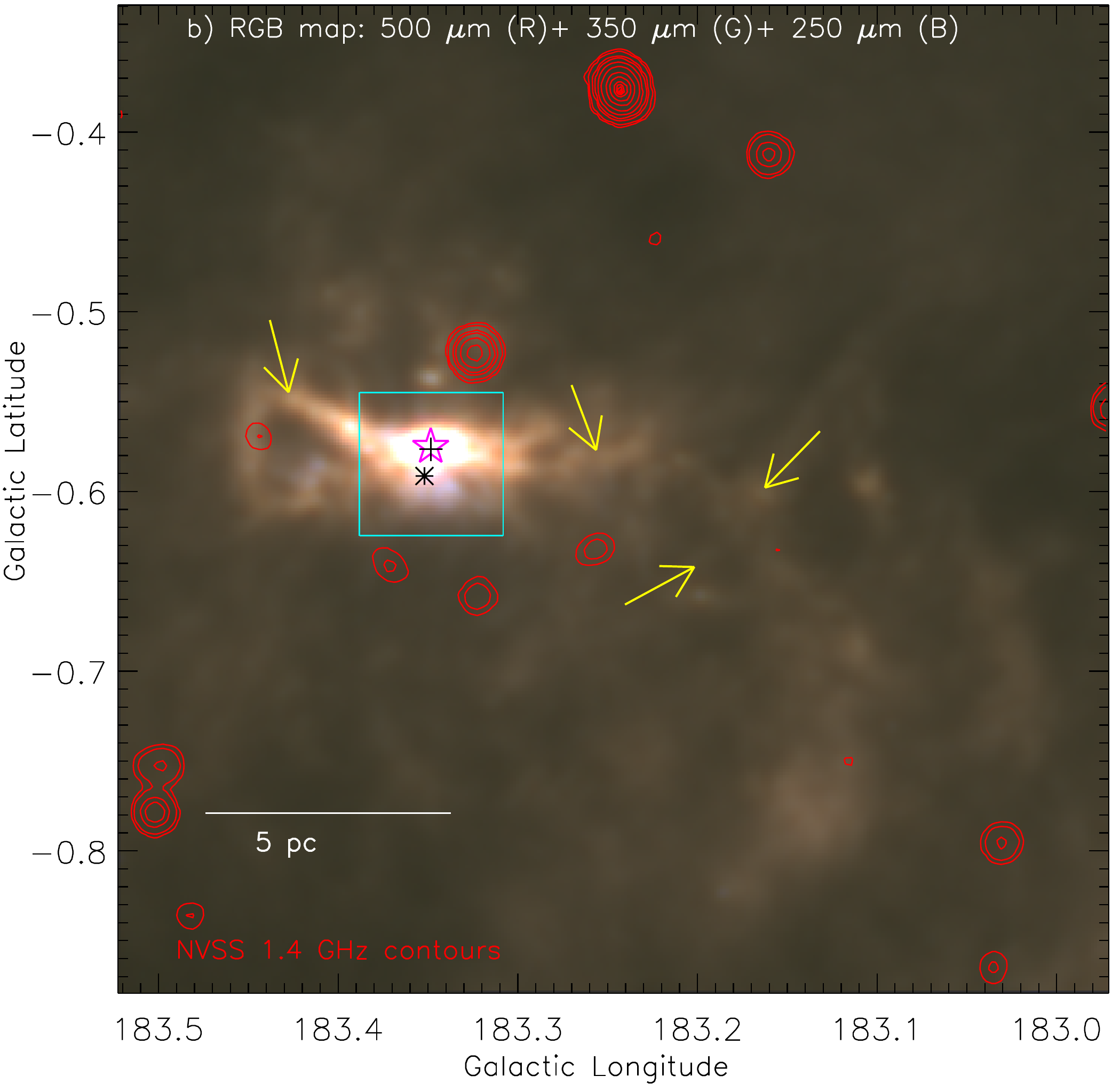}
\caption{\scriptsize Large-scale view of I05480+2545 (size of the selected field $\sim$0$\degr$.55 
$\times$ 0$\degr$.55 ($\sim$20.1 pc $\times$ 20.1 pc at a distance of 2.1 kpc); central coordinates: $l$ = 183$\degr$.247; $b$ = $-$0$\degr$.605). 
a) {\it WISE} 12 $\mu$m image of the region probed in this paper. 
The dotted-dashed red box encompasses the area shown in Figures~\ref{fs2}a 
and~\ref{fs2}b. b) A three color-composite map ({\it Herschel} 500 $\mu$m (red), 350 $\mu$m (green), and 250 $\mu$m (blue) images in log scale). 
The composite map is also overlaid with the NVSS 1.4 GHz contours. 
The 1.4 GHz contours (in red) are superimposed with 
levels of 0.5, 1,  3, 5, 10, 20, 40, 60, 80, 90, 95, and 98\% of the peak value (i.e.  0.437 Jy/beam).
The solid cyan box encompasses the area shown in Figure~\ref{fs3}. 
In both the panels, four arrows highlight the infrared dark regions. 
In each panel, the positions of I05480+2545 (+), IRAS 05480+2544 ($\ast$), and 6.7 GHz MME ($\star$) are marked.  
In both the panels, the scale bar corresponding to 5 pc (at a distance of 2.1 kpc) is shown in the bottom left corner.}
\label{fs1}
\end{figure*}

\begin{figure*}
\epsscale{0.61}
\plotone{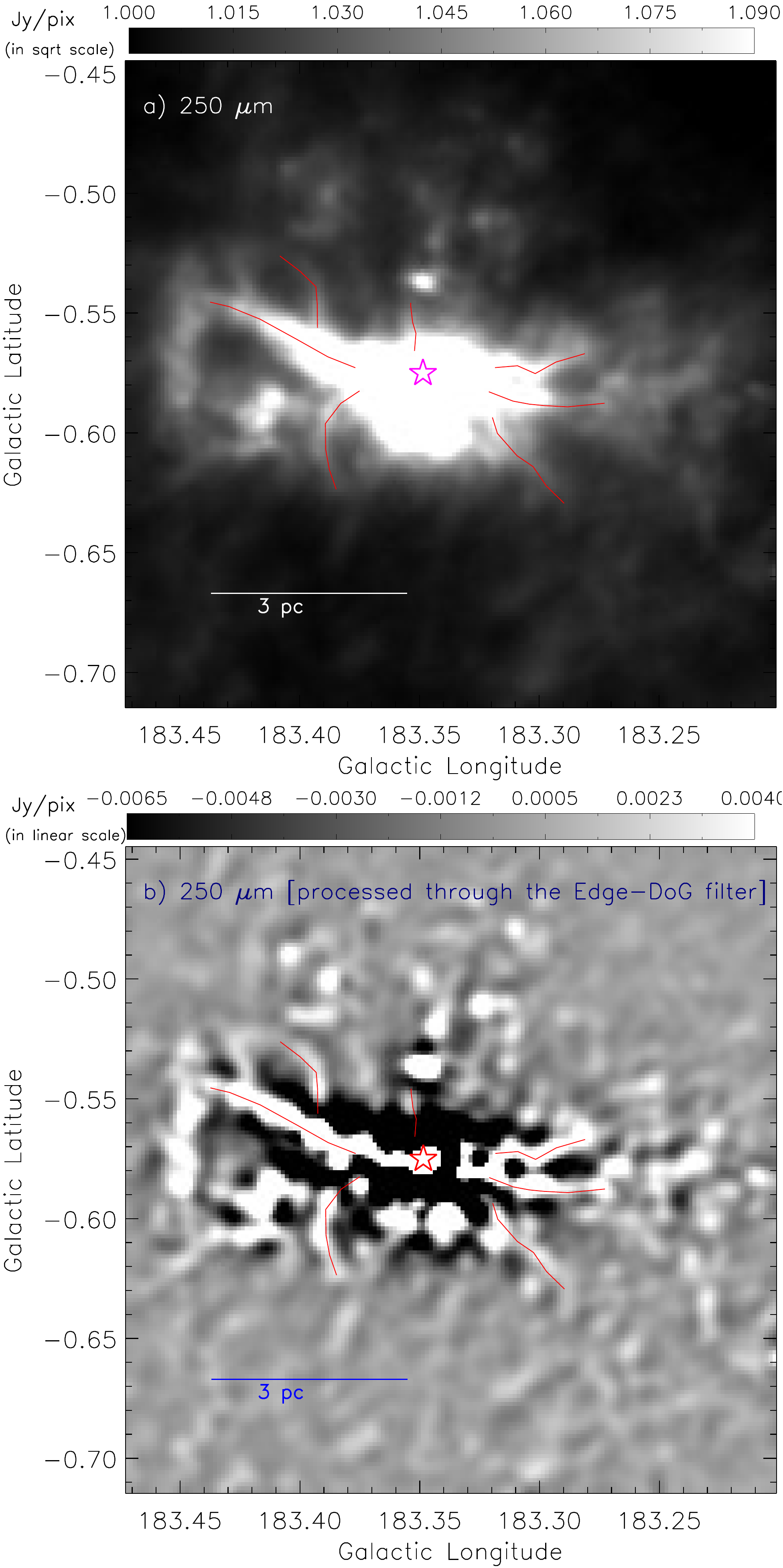}
\caption{\scriptsize The embedded {\it Herschel} filaments around I05480+2545. 
a) The filaments detected in the {\it Herschel} 250 $\mu$m image is highlighted on the {\it Herschel} 250 $\mu$m image.
(see solid red curves). b) The {\it Herschel} 250 $\mu$m image is processed through an ``Edge-DoG" algorithm.
The {\it Herschel} filaments are also highlighted by solid red curves.
The position of the 6.7 GHz MME is highlighted by a star in each panel. 
In both the panels, the scale bar corresponding to 3 pc (at a distance of 2.1 kpc) is 
shown in the bottom left corner.}
\label{fs2}
\end{figure*}

Figures~\ref{fs1}a and~\ref{fs1}b reveal the spatial locations of ionized, warm dust, and cold dust emissions in our selected site around I05480+2545 
(selected field $\sim$0$\degr$.55 $\times$ 0$\degr$.55).
In Figure~\ref{fs1}a, we show a gray-scale image at {\it WISE} 12 $\mu$m, indicating the presence of the warm dust emission. 
Figure~\ref{fs1}b shows a color-composite image produced using the {\it Herschel} 500 $\mu$m (red), 350 $\mu$m (green), and 250 $\mu$m (blue) images, 
tracing the cold dust emission (see Section~\ref{subsec:temp} for quantitative estimates). 
The 12 $\mu$m continuum image shows several dark regions which are highlighted by arrows in Figure~\ref{fs1}a. 
However, these dark regions appear as bright emission regions in the {\it Herschel} images at 250--500 $\mu$m (also see arrows in Figure~\ref{fs1}b). 
The {\it Herschel} composite image also reveals a bright condensation which harbors the 6.7 GHz MME. 
The noticeable emission observed in the {\it Herschel} images is well located within the MCI05, as previously reported in the $^{13}$CO intensity map 
\citep[see ID \#79 or BFS 48 region in Figure 9j in][]{kawamura98}. 
The NVSS 1.4 GHz emission is also superimposed on the {\it Herschel} composite image. 
We do not find any ionized emission toward the condensation containing the 6.7 GHz MME. 
In Figure~\ref{fs2}a, we have found embedded filaments (length $\sim$1--3 pc) in the {\it Herschel} 250 $\mu$m image. 
These filaments are identified based on a visual inspection of the {\it Herschel} image. We have also employed an edge detection algorithm 
\citep[i.e. Difference of Gaussian (DoG); see][]{gonzalez11,assirati14} to visually find these embedded filaments against the bright emission seen in the {\it Herschel} images. The technique consists of subtracting two Gaussian kernels, where a kernel has a standard deviation relatively lower than the previous one \citep[i.e.][]{assirati14}. In this work, we used two radius values (i.e.  3 and 5 pixels) of the Gaussians and the features detected by the DoG filter depend on the difference between these two radius values.
Figure~\ref{fs2}b shows the {\it Herschel} 250 $\mu$m image where a DoG filter has been applied. 
There are several filaments seen in the {\it Herschel} image. 
At least seven {\it Herschel} filaments are highlighted in the image and appear to be radially pointed to the condensation 
associated with the 6.7 GHz MME, forming a ``hub-filament" system. 

\begin{figure*}
\epsscale{1}
\plotone{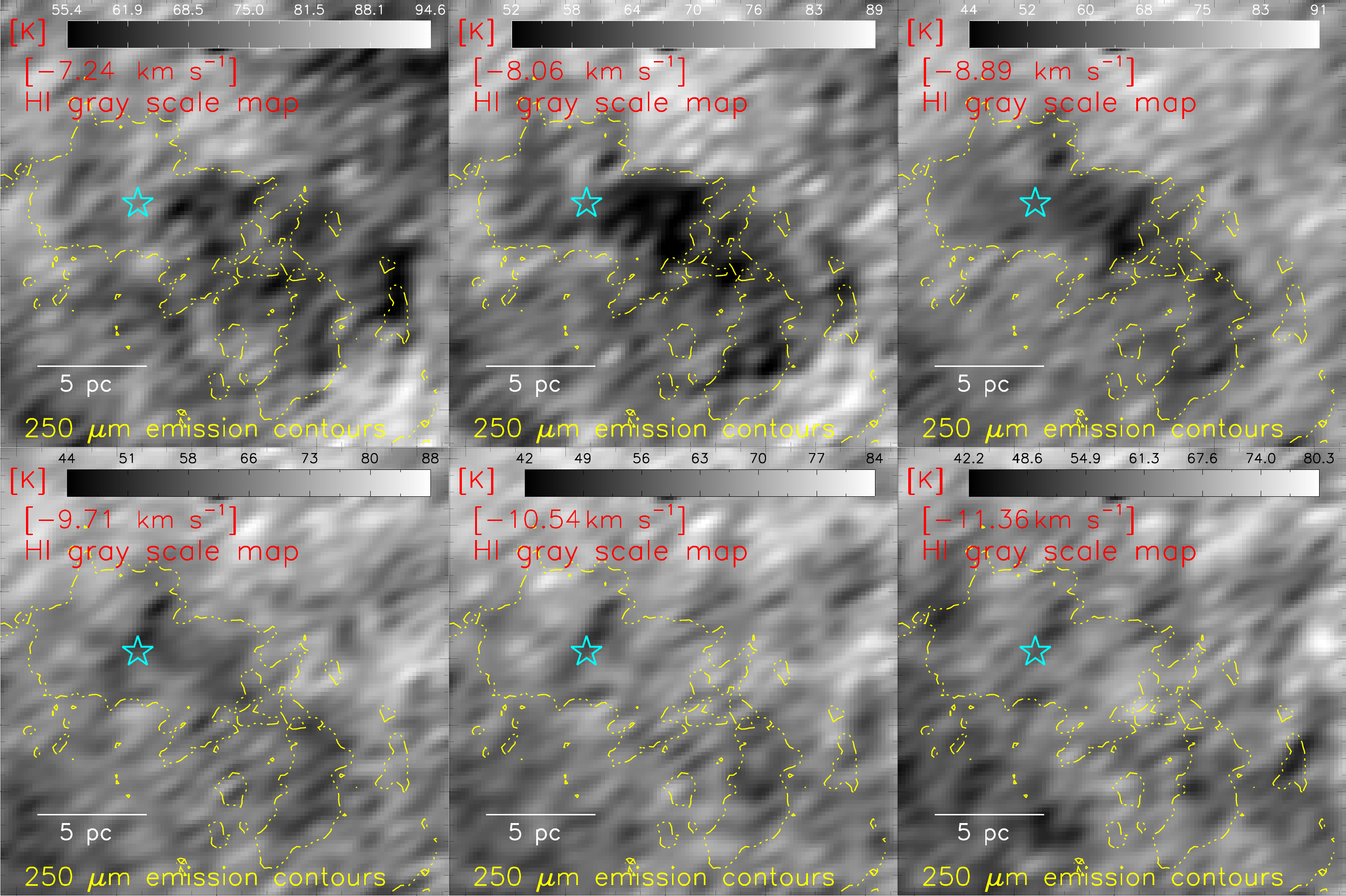}
\caption{\scriptsize CGPS 21 cm H\,{\sc i} velocity channel maps of I05480+2545 are 
overlaid with the {\it Herschel} 250 $\mu$m emission contour. 
The velocity value (in km s$^{-1}$) is labeled in each panel. 
The position of the 6.7 GHz MME is highlighted by a cyan star in all the panels. 
In each panel, the 250 $\mu$m continuum contour (in dotted-dashed yellow) is superimposed with a level of 0.072 Jy/pix. 
The H\,{\sc i} maps show H\,{\sc i} self-absorption features \citep[i.e., cold H\,{\sc i} traced in absorption against warmer background H\,{\sc i};][]{kerton05}.}
\label{fs5}
\end{figure*}

In Figure~\ref{fs5}, using the 21 cm H\,{\sc i} line data, we show velocity channel maps of 21 cm H\,{\sc i} (velocity range from $-$11.36 to $-$7.24 km s$^{-1}$) toward the MCI05, which is overlaid with the {\it Herschel} 250 $\mu$m emission contour. 
In the H\,{\sc i} channel maps, there are low intensity ($<$ 70 K) regions within the MCI05 seen as the black or dark gray regions. 
These regions trace the H\,{\sc i} self-absorption (HISA) features \citep[e.g.,][]{kerton05} 
that can indicate the presence of the residual amounts of very cold H\,{\sc i} gas in the molecular cloud.
\subsubsection{Multi-band picture of I05480+2545 and 6.7 GHz MME}
\label{sec:clmtv}
\begin{figure*}
\epsscale{0.87}
\plotone{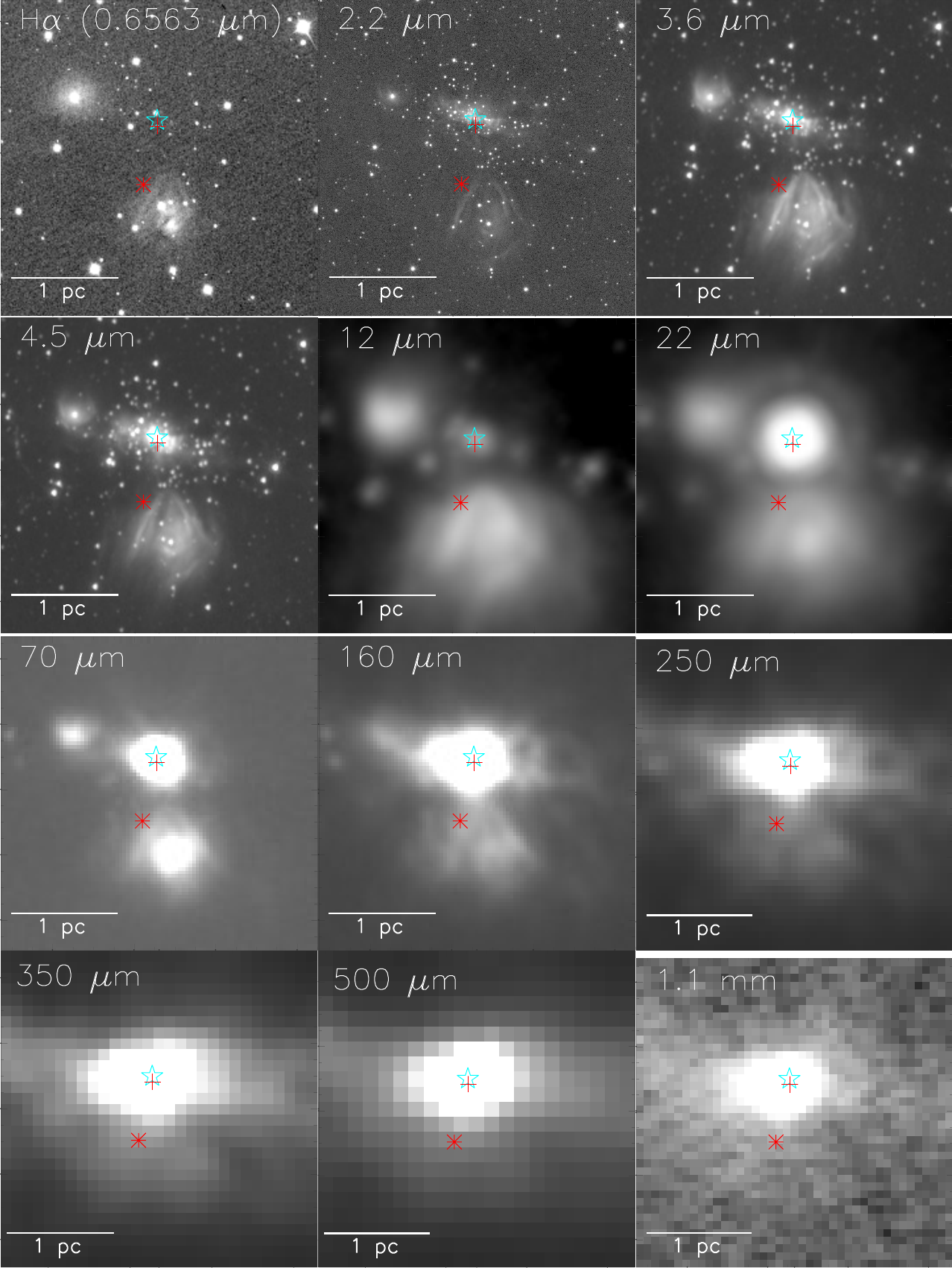}
\caption{\scriptsize A multi-wavelength view of I05480+2545 (size of the selected field $\sim$4$\farcm8$ $\times$ 4$\farcm8$ 
($\sim$2.9 pc $\times$ 2.9 pc at a distance of 2.1 kpc); central coordinates: $l$ = 183$\degr$.348; $b$ = $-$0$\degr$.585). 
The panels show images at 0.6563 $\mu$m (H$\alpha$; with continuum), 2.2 $\mu$m, 3.6 $\mu$m, 4.5 $\mu$m, 12 $\mu$m, 22 $\mu$m, 70 $\mu$m, 160 $\mu$m, 250 $\mu$m, 350 $\mu$m, 500 $\mu$m, 1.1 mm, from the IPHAS, UKIDSS-GPS, {\it Spitzer}, {\it Herschel}, and Bolocam (from left to right in increasing order). 
The position of the 6.7 GHz MME is highlighted by a cyan star in all the panels. 
In all the panels, other marked symbols are similar to those shown in Figure~\ref{fs1}b. 
In all the panels, the scale bar corresponding to 1 pc (at a distance of 2.1 kpc) is shown in the bottom left corner.}
\label{fs3}
\end{figure*}

In Figure~\ref{fs3}, we show images of I05480+2545 (selected field $\sim$4$\farcm8$ $\times$ 4$\farcm8$) covering from optical H$\alpha$, NIR, MIR, 
FIR, sub-mm, and millimeter wavelengths. The noticeable diffuse emissions are evident in the H$\alpha$ image and are away from the 6.7 GHz MME.
These diffuse emissions may trace the ionized emission. We also find several embedded infrared sources near the positions of the 6.7 GHz MME 
and I05480+2545. An almost inverted V-like feature located near IRAS 05480+2544 is also seen at wavelengths 
shorter than 250 $\mu$m and is also associated with the H$\alpha$ diffuse emission. 
However, the inverted V-like feature is more prominently seen in the images at 2.2--4.5 $\mu$m. 
The condensation containing the 6.7 GHz MME is also observed in the Bolocam map at 1.1 mm.

\begin{figure*}
\epsscale{0.475}
\plotone{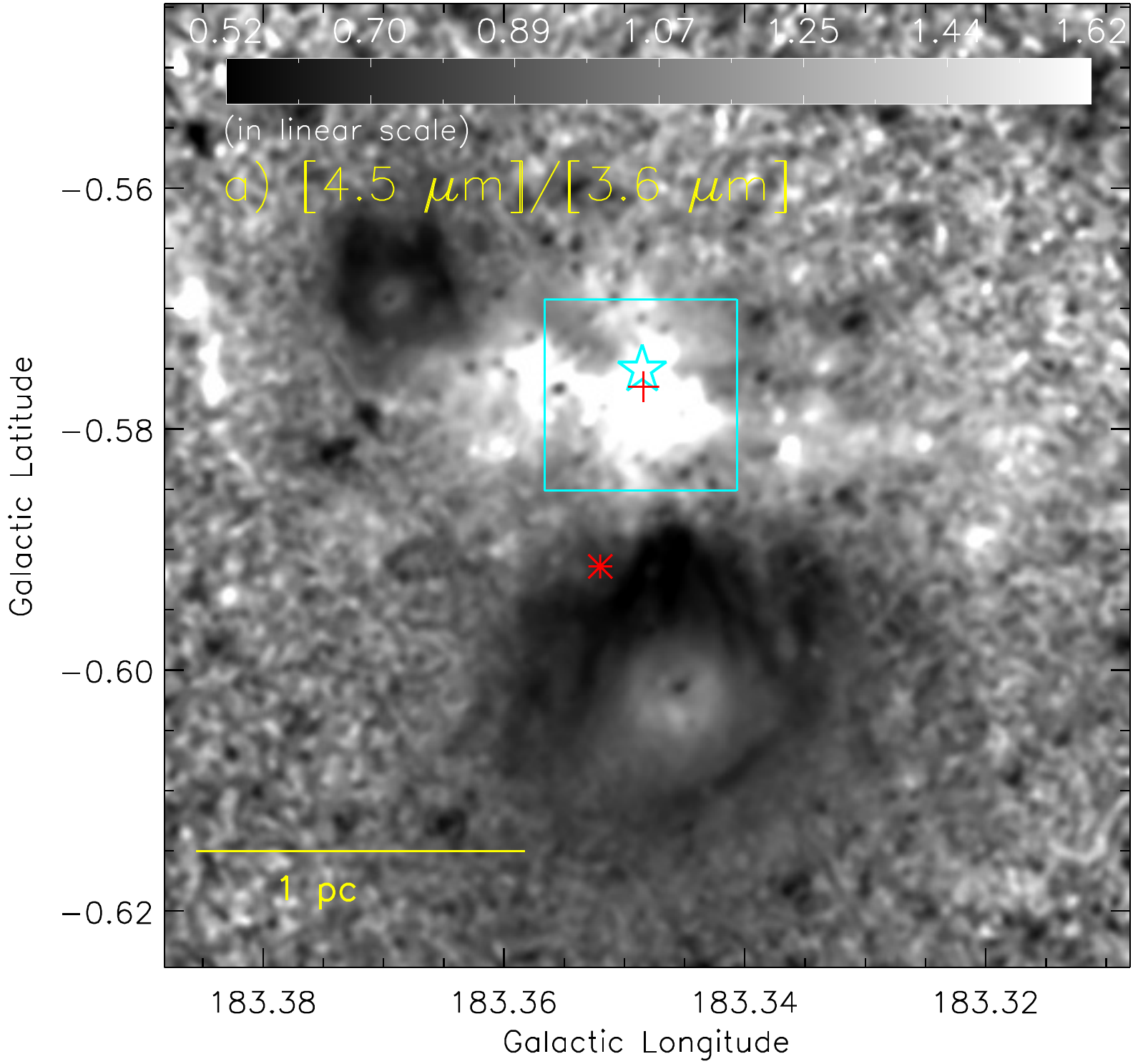}
\epsscale{0.455}
\plotone{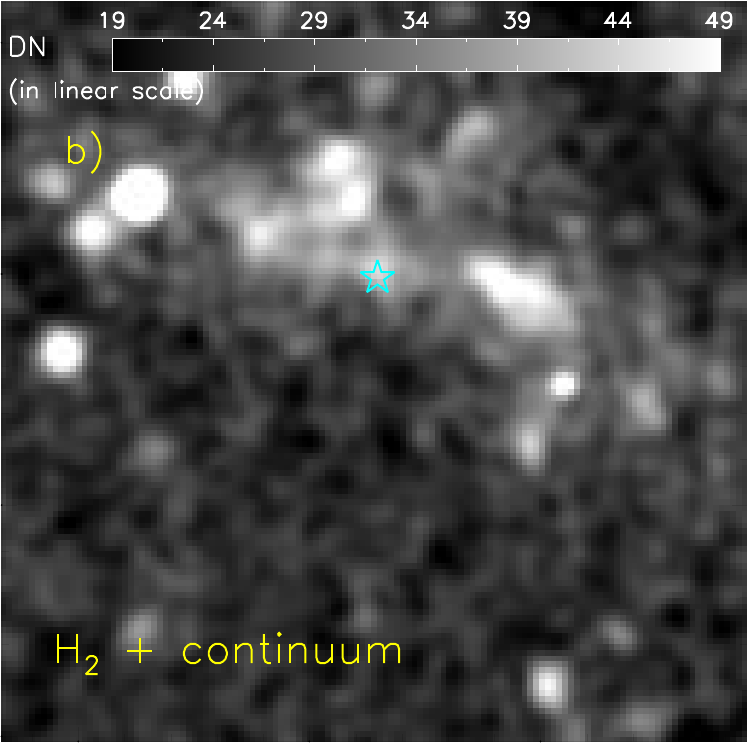}
\epsscale{0.455}
\plotone{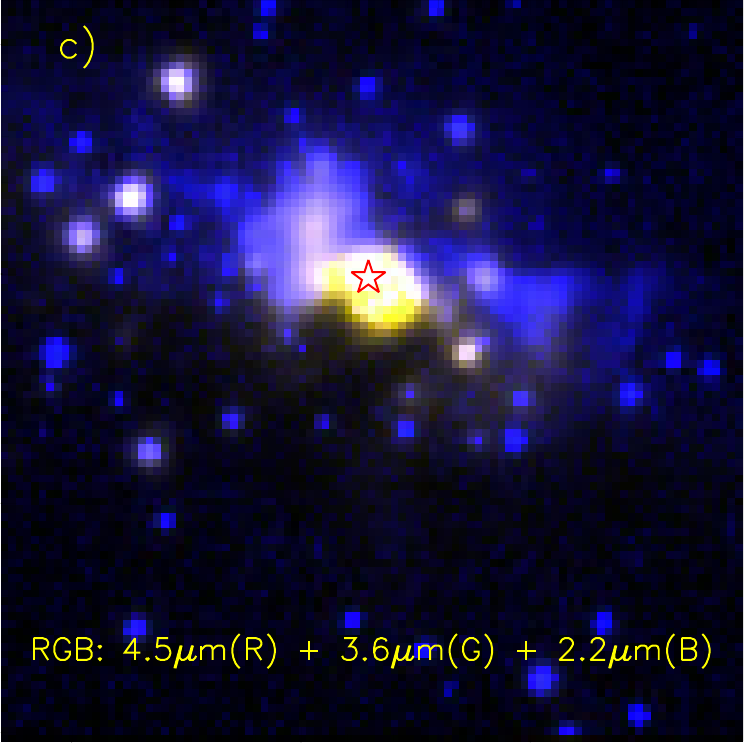}
\epsscale{0.455}
\plotone{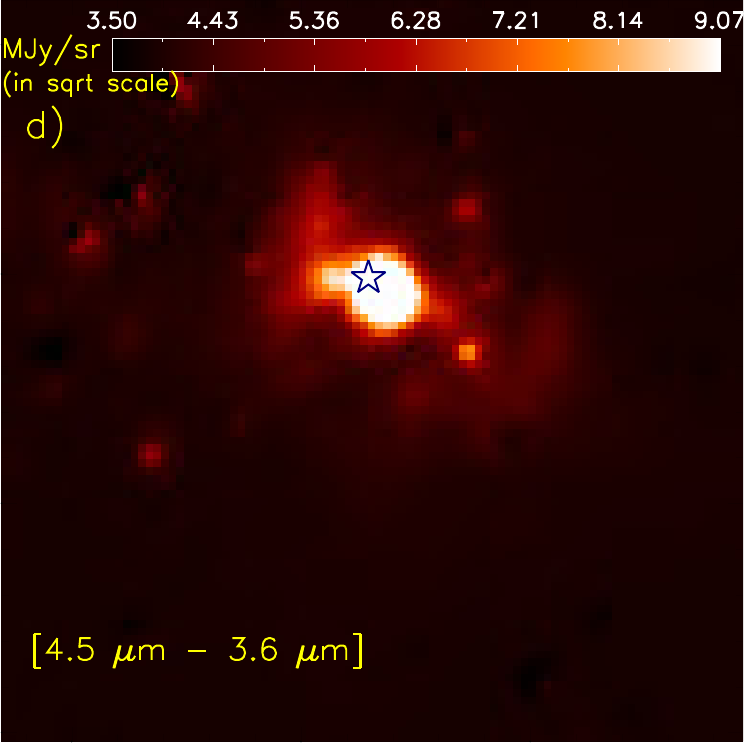}
\caption{\scriptsize a) {\it Spitzer} ratio map of 4.5 $\mu$m/3.6 $\mu$m emission. 
The ratio map is exposed to median filtering with a width of 5 pixels and smoothened by 
4 pixel $\times$ 4 pixel using a boxcar algorithm. The solid cyan box encompasses the area shown in 
Figures~\ref{fs4}b,~\ref{fs4}c, and~\ref{fs4}d (size of the selected field $\sim$58$\arcsec$ $\times$ 58$\arcsec$; 
central coordinates: $l$ = 183$\degr$.348; $b$ = $-$0$\degr$.577). Other marked symbols are similar to those shown in Figure~\ref{fs1}b and 
the scale bar corresponding to 1 pc (at a distance of 2.1 kpc) is shown in the bottom left corner. 
 b) The map shows the central region in zoomed-in view, using 
the MIRO H$_{2}$ image (with continuum) (gray scale; see a solid cyan box in Figure~\ref{fs4}a). 
c) A zoomed-in view of I05480+2545 using a three-color composite image ({\it Spitzer} 4.5 $\mu$m (in red), 
{\it Spitzer} 3.6 $\mu$m (in green), and UKIDSS-GPS 2.2 $\mu$m (in blue)). 
d) A false color {\it Spitzer} difference map of 4.5 $\mu$m $-$ 3.6 $\mu$m emission. 
The position of the 6.7 GHz MME is highlighted by a star in all the panels.}
\label{fs4}
\end{figure*}

Following the analysis presented in \citet{dewangan16,dewangan17a}, the {\it Spitzer}-IRAC ratio map of 4.5 $\mu$m/3.6 $\mu$m emission is examined 
to infer the signatures of molecular outflows and the impact of massive stars on their surroundings. It is known that the IRAC 3.6 $\mu$m and 4.5 $\mu$m bands have nearly exact point response functions (PRFs), allowing to directly use the ratio of 4.5 $\mu$m to 3.6 $\mu$m images.
We present the {\it Spitzer}-IRAC ratio map of I05480+2545 in Figure~\ref{fs4}a. 
In Figure~\ref{fs4}a, we find the dark/black regions in the ratio map of 4.5 $\mu$m/3.6 $\mu$m emission, 
tracing the excess of 3.6 $\mu$m emission. These dark regions are spatially correlated with the 
diffuse H$\alpha$ emission (see panel of the H$\alpha$ image in Figure~\ref{fs3}). 
In the ratio map of 4.5 $\mu$m/3.6 $\mu$m emission, we also detect the bright regions near the 6.7 GHz MME, 
depicting the excess of 4.5 $\mu$m emission. 
IRAC 3.6 $\mu$m band contains polycyclic aromatic hydrocarbon (PAH) emission 
at 3.3 $\mu$m as well as a prominent molecular hydrogen feature at 3.234 $\mu$m ($\nu$ = 1--0 $ O$(5)). 
IRAC 4.5 $\mu$m band harbors a hydrogen recombination line Br$\alpha$ (4.05 $\mu$m) and 
a prominent molecular hydrogen line emission ($\nu$ = 0--0 $S$(9); 4.693 $\mu$m), which is excited by outflow shocks.
Taking into account the presence of PAH feature (at 3.3 $\mu$m) in the 3.6 $\mu$m band, the dark regions surrounding 
the ionized emission, as traced in the H$\alpha$ image, appear to be produced due to the impact of ionizing photons. 
In the ratio map, we find two such dark regions (i.e. a V-like feature linked with the BFS 48 site and a region in the galactic north-east direction with respect to the 6.7 GHz MME). 
Note that the NVSS 1.4 GHz emission is absent toward both the dark regions, indicating that the sensitivity of 
the 1.4 GHz data is not enough to detect radio continuum emission toward these dark regions. 
Furthermore, the bright emission regions near the 6.7 GHz MME, where the radio continuum emission is absent and embedded NIR sources are found, appear to probably trace the outflow activity. This interpretation can be supported by the presence of the diffuse H$_{2}$ emission 
(without continuum subtracted) in the MIRO H$_{2}$ map (see Figure~\ref{fs4}b). 
In Section~\ref{subsec:phot1}, we have carried out the quantitative analysis of the infrared excess sources.

Figures~\ref{fs4}b,~\ref{fs4}c, and~\ref{fs4}d show a zoomed-in view of the 6.7 GHz MME. 
Figure~\ref{fs4}b shows the MIRO H$_{2}$ map of the 6.7 GHz MME. 
In Figure~\ref{fs4}c, we show a color-composite image using the UKIDSS-GPS K and {\it Spitzer} 3.6--4.5 $\mu$m images. 
An infrared counterpart (IRc) of the 6.7 GHz MME is found at wavelengths longer than 2 $\mu$m. 
Additionally, the extended emission traced in the K and {\it Spitzer} images is linked with the IRc that does not look like a point-like source. 
The K band contains H$_{2}$ (2.12 $\mu$m) and Br$\gamma$ (2.166 $\mu$m) lines. 
However, there is no radio continuum emission seen toward the IRc, suggesting the detection of Br$\gamma$ emission is unlikely. 
Hence, it seems that there is detection of diffuse H$_{2}$ emission 
(also see H$_{2}$ map in Figure~\ref{fs4}b), suggesting the presence of the outflow activity. 
In Figure~\ref{fs4}d, we show a {\it Spitzer}-IRAC difference map of 4.5 $\mu$m$-$3.6 $\mu$m emission, 
tracing the similar diffuse emission as seen in the K-band image.
Hence, it seems that the IRc of the 6.7 GHz MME could be a MYSO candidate associated with the molecular outflow, prior to the UCH\,{\sc ii} phase.  
This argument is supported by the exclusive association of Class~II 6.7 GHz methanol masers with MYSOs \citep[e.g.][]{urquhart13}. 
Furthermore, in the W42 and IRAS 17599-2148 sites, the IRc of the 6.7 GHz MME has also been identified and characterized as a MYSO \citep{dewangan15a,dewangan16b}.

However, our initial results can be further explored using high-resolution data at longer wavelengths to investigate the physical properties of 
IRc and its formation process. 
\subsubsection{{\it Herschel} temperature and column density maps}
\label{subsec:temp}
In this section, the {\it Herschel} temperature and column density maps of MCI05 are presented. 
Following the methods given in \citet{mallick15}, these maps are generated from a 
pixel-by-pixel spectral energy distribution (SED) fit with a modified blackbody to the cold dust emission at {\it Herschel} 160--500 $\mu$m \citep[also see][]{dewangan15}. The {\it Herschel} 70 $\mu$m data are excluded from the analysis, because the 70 $\mu$m emission is 
dominated by the ultraviolet (UV) heated warm dust. In the following, we provide a brief step-by-step explanation of the adopted methods. 

The {\it Herschel} images at 250--500 $\mu$m are in units of surface brightness, MJy sr$^{-1}$, while the image at 160 $\mu$m is in unit of Jy pixel$^{-1}$.
The plate scales of 160, 250, 350, and 500 $\mu$m images are 3$''$.2, 6$''$, 10$''$, 
and 14$''$ pixel$^{-1}$, respectively.  
Before the SED fit, the 160--350 $\mu$m images were convolved to the lowest angular 
resolution of 500 $\mu$m image ($\sim$37$\arcsec$) 
and were converted into the same flux unit (i.e. Jy pixel$^{-1}$). 
Furthermore, we regridded these images to the pixel size of 500 $\mu$m image ($\sim$14$\arcsec$). 
These steps were performed using the convolution kernels available in the HIPE software. 
Next, the sky background flux level was estimated to be 0.060, 0.133, 0.198, and $-$0.095 Jy pixel$^{-1}$ for the 500, 350, 250, and 
160 $\mu$m images (size of the selected region $\sim$13$\farcm$4 $\times$ 14$\farcm$8; 
centered at:  $l$ = 183$\degr$.181; $b$ = $-$0$\degr$.354), respectively. 
To avoid diffuse emission linked with the selected target, the featureless dark field away from the MCI05 was 
carefully selected for the background estimation. 

\begin{figure*}
\epsscale{0.375}
\plotone{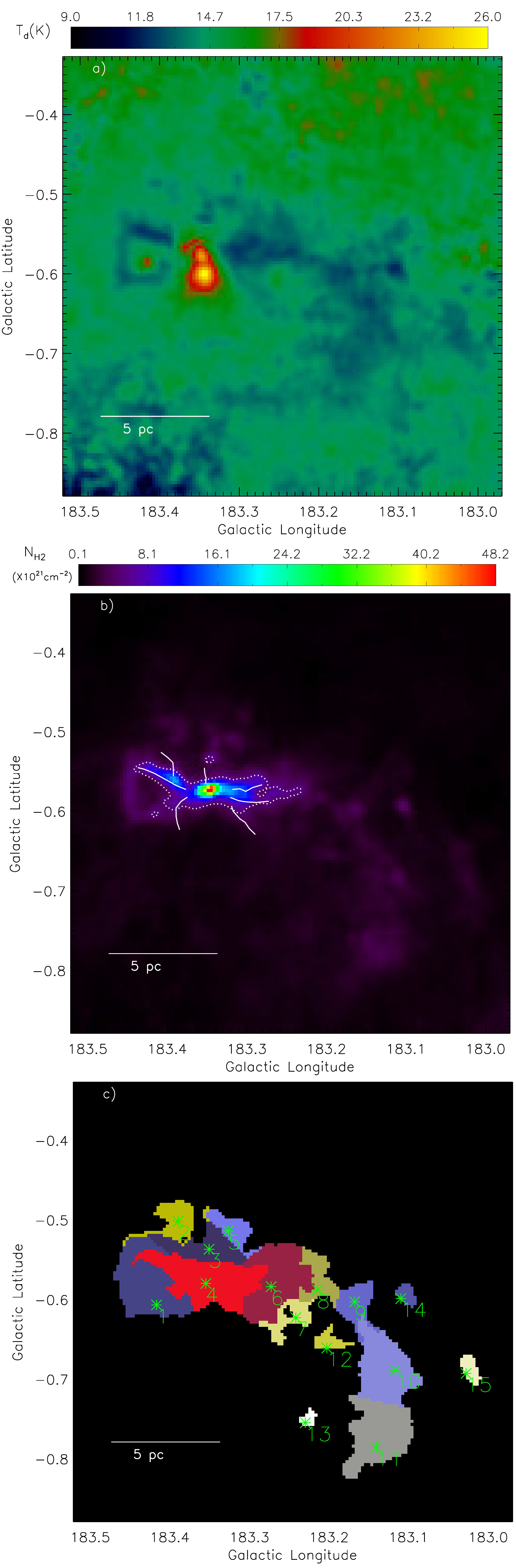}
\caption{\scriptsize a) {\it Herschel} temperature map of I05480+2545.
b) {\it Herschel} column density ($N(\mathrm H_2)$) map of I05480+2545. 
The map is overlaid with a column density contour (see dotted white contour). 
The column density map can be used to infer the extinction with $A_V=1.07 \times 10^{-21}~N(\mathrm H_2)$ \citep{bohlin78}. 
A contour level is shown at 5.2 $\times$ 10$^{21}$ cm$^{-2}$ (A$_{V}$ $\sim$5.5 mag). 
The {\it Herschel} filaments are also highlighted by solid white curves (see Figure~\ref{fs2}a). 
c) Based on the {\it Herschel} column density map, the identified clumps are 
marked by asterisk symbols and the boundary of each {\it Herschel} clump is also shown in the figure. 
The boundary of each {\it Herschel} clump is highlighted along with its corresponding clump ID (see Table~\ref{tab1}).
In all the panels, the scale bar corresponding to 5 pc (at a distance of 2.1 kpc) is shown in the bottom left corner. }
\label{fs6}
\end{figure*}

Finally, to produce the temperature and column density maps, a modified blackbody was fitted to the observed fluxes on a pixel-by-pixel basis 
\citep[see equations 8 and 9 in][]{mallick15}. 
The fitting was carried out using the four data points for each pixel, maintaining the 
dust temperature (T$_{d}$) and the column density ($N(\mathrm H_2)$) as free parameters. 
In the analysis, we adopted a mean molecular weight per hydrogen molecule ($\mu_{H2}$) of 2.8 
\citep{kauffmann08} and an absorption coefficient ($\kappa_\nu$) of 0.1~$(\nu/1000~{\rm GHz})^{\beta}$ cm$^{2}$ g$^{-1}$, 
including a gas-to-dust ratio ($R_t$) of 100, with a dust spectral index ($\beta$) of 2 \citep[see][]{hildebrand83}. 
The final temperature and column density maps (resolution $\sim$37$\arcsec$) are shown in Figures~\ref{fs6}a and~\ref{fs6}b, respectively.
In the {\it Herschel} temperature map, the condensation containing the 6.7 GHz MME and V-like feature is traced in a temperature (T$_{d}$) 
range of $\sim$18-26 K (see Figure~\ref{fs6}a). The areas containing the HISA features are depicted in a temperature range of about 10--12~K. 
Using the NH$_{3}$ line data, \citet{wu10} computed the physical parameters of dense gas (i.e. the gas kinetic temperature 
and rotational temperature) toward the condensation containing the 6.7 GHz MME. 
The kinetic temperature of the MME condensation was reported to be $\sim$20 K, which is in agreement with our estimated temperature value. 
The condensation containing the 6.7 GHz MME is located in the highest column density region with 
peak $N(\mathrm H_2)$ $\sim$4.8~$\times$~10$^{22}$ cm$^{-2}$ corresponding to A$_{V}$ $\sim$51 mag (see Figure~\ref{fs6}b). 
Here, the conversion is carried out using a relation between optical extinction and hydrogen column density 
\citep[i.e. $A_V=1.07 \times 10^{-21}~N(\mathrm H_2)$;][]{bohlin78}. 
\subsubsection{{\it Herschel} clumps}
\label{sec:clmv}
The column density map can also be used to find embedded clumps in a given star-forming regions. 
In the column density map (see Figure~\ref{fs6}b), the ``{\it clumpfind}" IDL program \citep{williams94} has been used to identify the clumps and 
to compute their total column densities. 
Fifteen clumps are found in the map and are labeled in Figure~\ref{fs6}c. 
Furthermore, the boundary of each clump is also highlighted in Figure~\ref{fs6}c. 
We have also estimated the mass of each clump using its total column density. 
The mass of a single {\it Herschel} clump is determined using the following formula:
\begin{equation}
M_{clump} = \mu_{H_2} m_H Area_{pix} \Sigma N(H_2)
\end{equation}
where $\mu_{H_2}$ is assumed to be 2.8, $Area_{pix}$ is the area subtended by one pixel, and 
$\Sigma N(\mathrm H_2)$ is the total column density. 
The mass of each {\it Herschel} clump is listed in Table~\ref{tab1}. 
The table also gives an effective radius of each clump, which is an outcome of the {\it clumpfind} algorithm. 
It is found that the clump masses vary between 15 M$_{\odot}$ and 1875 M$_{\odot}$. 
The most massive clump (no 4, M$_{clump}$ $\sim$1875 M$_{\odot}$) contains the 6.7 GHz MME, 
where several filaments (see Figure~\ref{fs6}b) appear to be radially directed to this clump, revealing a ``hub-filament" morphology \citep[e.g.][]{myers09}. 
The implication of this configuration concerning the star formation process is discussed in Section~\ref{sec:disc}.
\subsection{Young stellar populations in the MCI05}
\label{subsec:phot1}
In this section, we describe the identification and classification procedures of YSOs using the 2MASS-GPS and GLIMPSE360 photometric data from 1--5 $\mu$m. 

\begin{figure*}
\epsscale{0.78}
\plotone{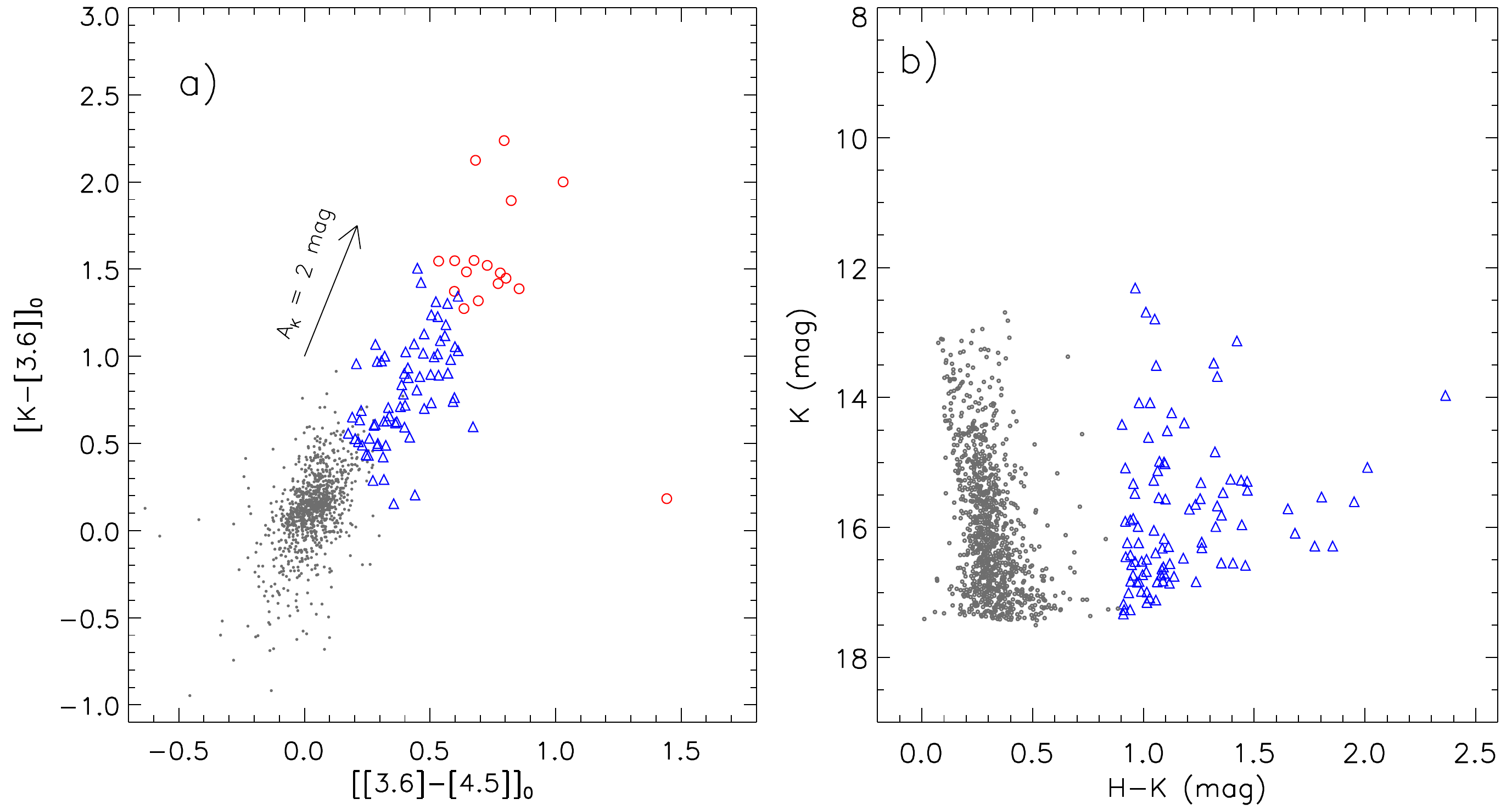}
\epsscale{0.5}
\plotone{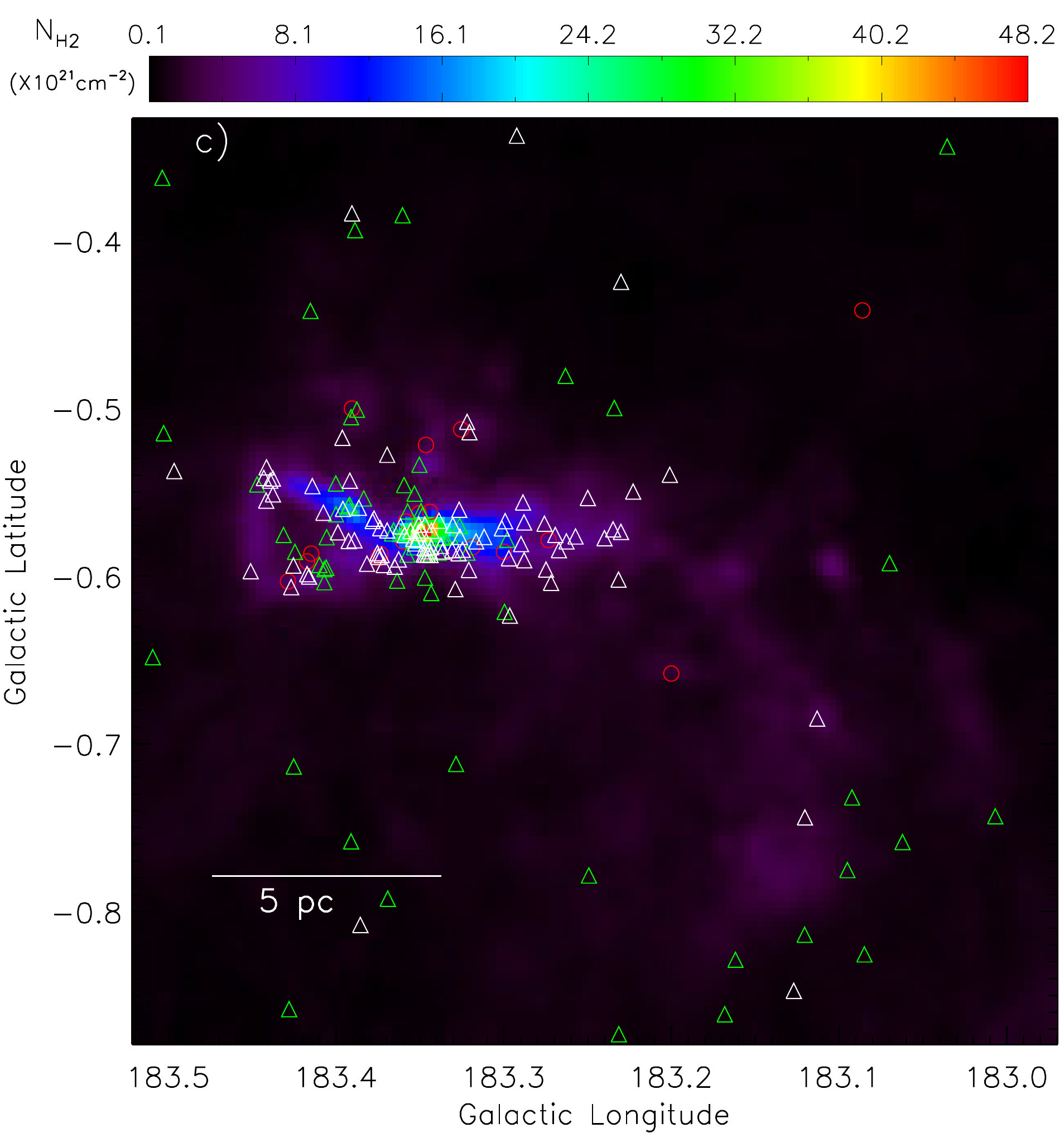}
\caption{\scriptsize The color-color and color-magnitude plots of point-like sources detected in our selected field around I05480+2545 (see Figure~\ref{fs1}).
a) The dereddened [K$-$[3.6]]$_{0}$ $vs$ [[3.6]$-$[4.5]]$_{0}$ color-color plot using the H, K, 3.6 $\mu$m, and 4.5 $\mu$m data (see text for details). 
In the plot, Class~I and Class~II YSOs are shown by red circles and open blue triangles, respectively. 
The extinction vector is shown using the average extinction laws from \citet{flaherty07}. 
b) Color-magnitude plot (H$-$K/K) of the sources observed only in H and K bands that have no counterparts 
in our selected GLIMPSE360 catalog. In the plot, Class~II YSOs are marked by open blue triangles. 
In each panel, the dots (in gray) show the stars with only photospheric emissions. 
Due to large fractions of stars with photospheric emissions, we have randomly shown only some of these stars in both the plots. 
In the color-color plot, we have shown only 1001 out of 3747 stars with photospheric emissions. 
In the color-magnitude plot, we have shown only 1001 out of 11738 stars with photospheric emissions. 
The positions of selected YSOs are marked in Figure~\ref{fs8}c.
c) The positions of YSOs are overlaid on the {\it Herschel} column density map.
The positions of Class~I and Class~II YSOs are marked by circles and triangles, respectively. 
The YSOs extracted using the H, K, 3.6 $\mu$m, and 4.5 $\mu$m data (see Figure~\ref{fs8}a) 
are highlighted by red circles and green triangles, whereas the white triangles refer the YSOs identified 
using the H and K bands (see Figure~\ref{fs8}b).}
\label{fs8}
\end{figure*}

The infrared excess emission of point-like sources is a very powerful utility to trace the embedded young stellar populations. 
The dereddened color-color space ([K$-$[3.6]]$_{0}$ and [[3.6]$-$[4.5]]$_{0}$) allows to identify infrared-excess sources \citep[e.g.][]{gutermuth09}. 
We obtained the dereddened color-color plot ([K$-$[3.6]]$_{0}$ and [[3.6]$-$[4.5]]$_{0}$) 
using the 2MASS and GLIMPSE360 photometric data at 1--5 $\mu$m. 
The dereddened colors were computed using the color excess ratios listed in \citet{flaherty07}. 
Based on the dereddened color conditions given in \citet{gutermuth09}, 
we select 86 (17 Class~I and 69 Class~II) YSOs in MCI05. 
One can also remove possible dim extragalactic contaminants from the selected YSOs with additional 
conditions (i.e., [3.6]$_{0}$ $<$ 15 mag for Class~I and [3.6]$_{0}$ $<$ 14.5 mag for Class~II) \citep[e.g.,][]{gutermuth09}.  
The dereddened 3.6 $\mu$m magnitudes were estimated using the observed color and the reddening laws \citep[from][]{flaherty07}.
In Figure~\ref{fs8}a, we present the color-color plot ([K$-$[3.6]]$_{0}$ versus [[3.6]$-$[4.5]]$_{0}$) for the sources. 
The selected Class~I and Class~II YSOs are marked by the red circles and blue triangles, respectively.\\  
Using the 2MASS-GPS data, we have also constructed a color-magnitude plot (H$-$K/K) to obtain additional young 
stellar populations in the MCI05 (see Figure~\ref{fs8}b). It is possible because the UKIDSS-GPS NIR data are three magnitudes deeper than 2MASS. 
The plot refers the red sources having H$-$K color $>$ 0.9 mag. 
This color criterion was selected based on the color-magnitude analysis of a nearby control field (selected size $\sim$0$\degr$.1 
$\times$ 0$\degr$.1; central coordinates: $l$ = 183$\degr$.536; $b$ = $-$0$\degr$.448). 
We obtain 95 additional YSOs using the color-magnitude space in the MCI05.\\ 

Together, we have selected a total of 181 YSOs in the MCI05 through the analysis of the 2MASS-GPS and GLIMPSE360 data at 1--5 $\mu$m. 
Figure~\ref{fs8}c shows the {\it Herschel} column density map overlaid with the positions of all the selected YSOs.
A majority of YSOs are seen toward the areas of high column density.
\subsection{Clustering of YSOs}
\label{subsec:srf}
\begin{figure*}
\epsscale{0.56}
\plotone{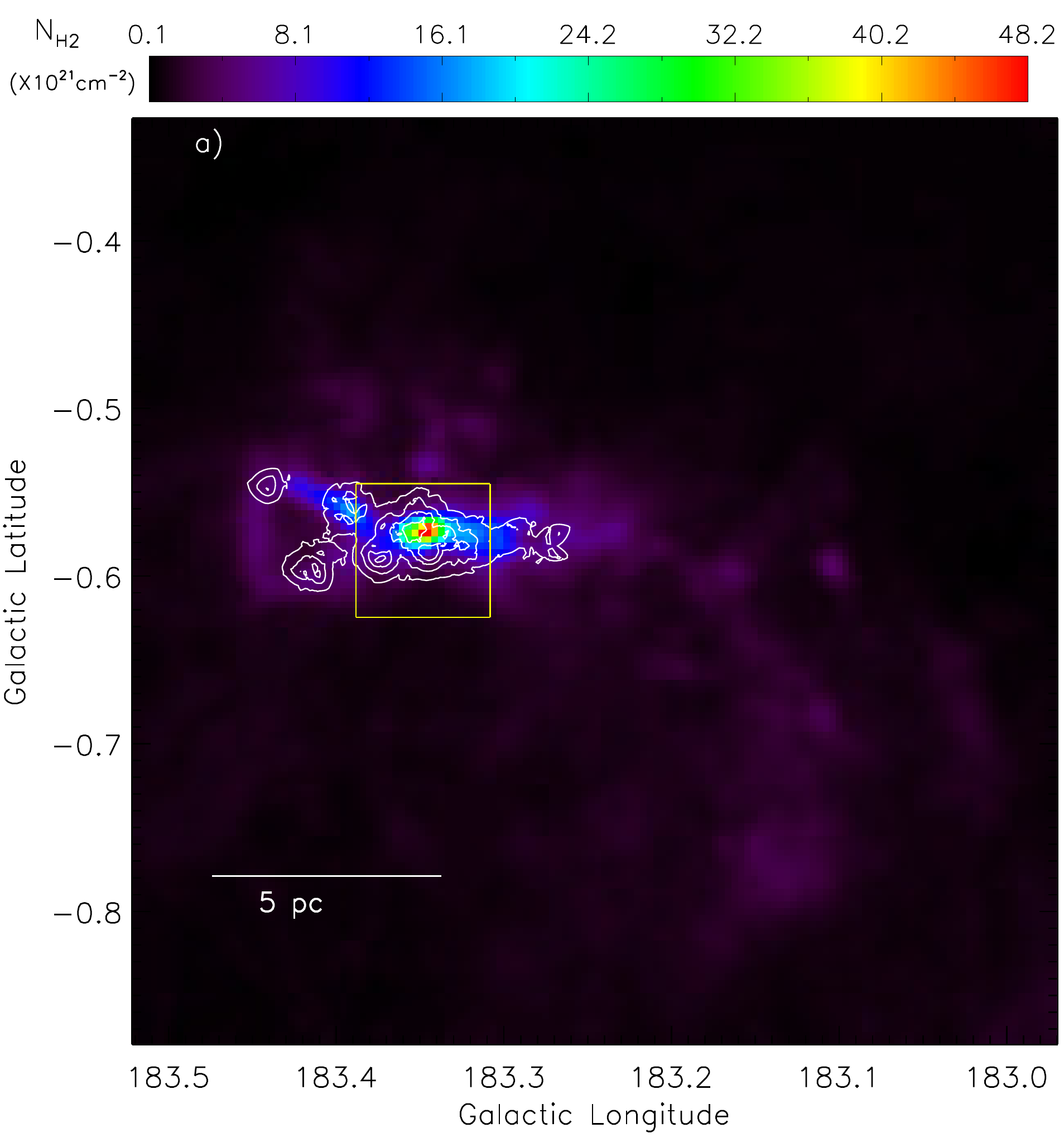}
\epsscale{0.56}
\plotone{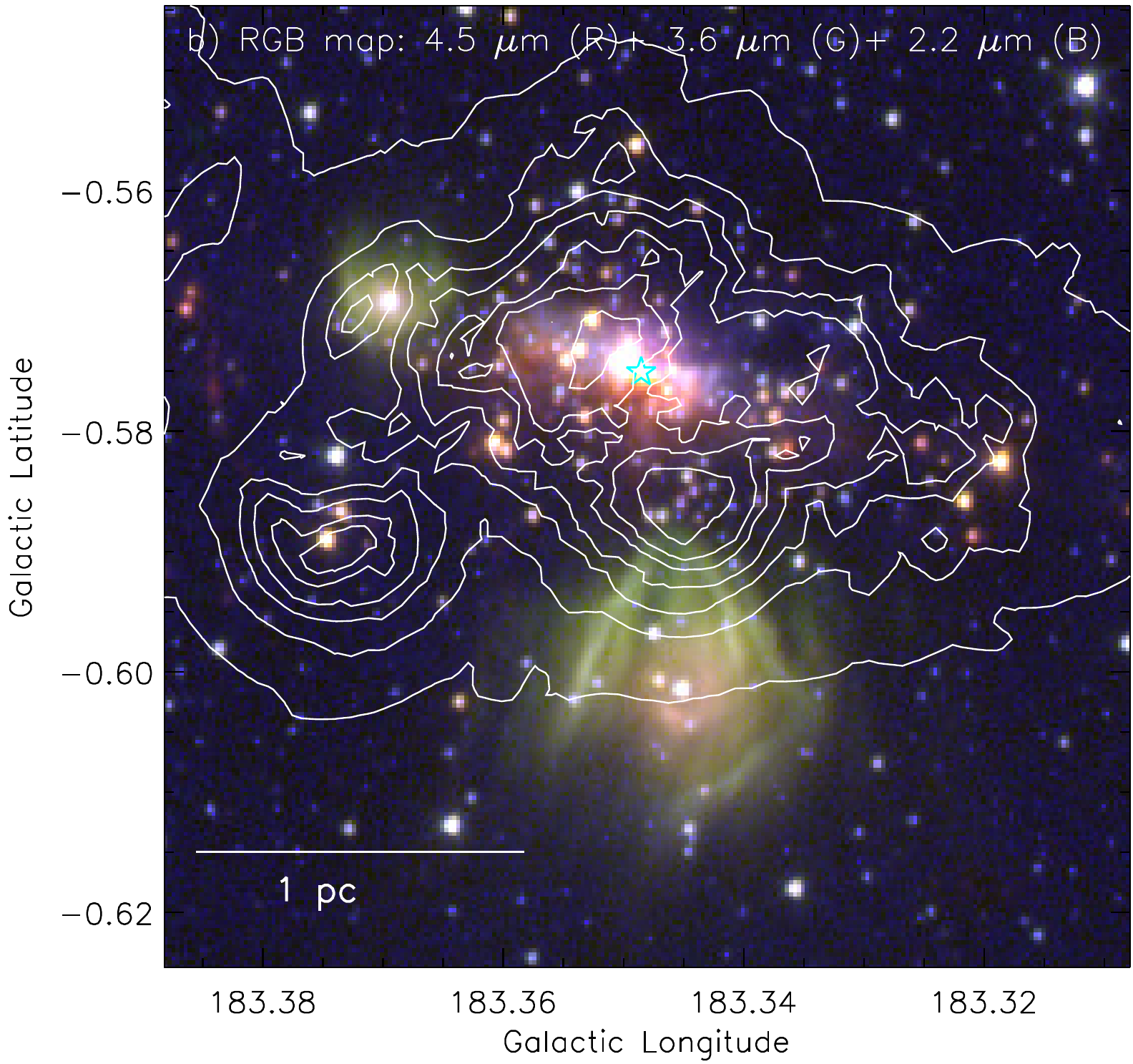}
\caption{\scriptsize a) Surface density contours (in white) of YSOs are superimposed 
on the {\it Herschel} column density map. 
The contours are shown at 5, 10, 20, and 40 YSOs/pc$^{2}$, from the outer to the inner side. 
The solid yellow box encompasses the area shown in Figure~\ref{fs9}b. 
The {\it Herschel} column density map is similar to the one shown in Figure~\ref{fs6}b. 
b) A color-composite map shows the central region (in zoomed-in view) overlaid with the 
surface density contours (see a solid yellow box in Figure~\ref{fs9}a). 
The map is produced using {\it Spitzer} 4.5 $\mu$m (red), {\it Spitzer} 3.6 $\mu$m (green), and UKIDSS-GPS 2.2 $\mu$m (blue) images. 
The contours are shown at 5, 10, 15, 20, 30, 40, and 70 YSOs/pc$^{2}$, from the outer to the inner side. 
The position of the 6.7 GHz MME is marked by a star.} 
\label{fs9}
\end{figure*}

To investigate the clustering of YSOs in the MCI05, the surface density map of YSOs is 
computed using the nearest-neighbour (NN) technique \citep[also see][for more details]{gutermuth09,bressert10}. 
We have produced the surface density map of the selected 181 YSOs, in a manner similar to that used in \citet{dewangan15}. 
The map is generated using a 5$\arcsec$ grid and 6 NN at a distance of 2.1 kpc. 
Figure~\ref{fs9}a shows the resultant surface density contours of YSOs overlaid on the {\it Herschel} column density map. 
The contour levels are shown at 5, 10, 20, and 40 YSOs/pc$^{2}$, increasing from the outer to the inner regions. 
The clusters of YSOs are found in the areas of high column density. 
Figure~\ref{fs9}b shows the surface density contours of YSOs overlaid on a color-composite image, using the 
GPS K and {\it Spitzer} images. 
The contours are shown at 5, 10, 15, 20, 30, 40, and 70 YSOs/pc$^{2}$, from the outer to the inner side. 
A large fraction of YSOs as well as their clustering are seen toward the most massive clump. 
This particular result helps us to trace the intense star formation activities toward the most massive clump. 
\section{Discussion}
\label{sec:disc}
In recent years, the {\it Herschel} continuum images have been used as an indispensable tool in the study of 
embedded filaments, and star-forming clumps in such system. 
In our selected site around I05480+2545, using the $^{13}$CO line data, 
an almost elongated filamentary cloud was observed by \citet{kawamura98}. 
A similar morphology is also found in the {\it Herschel} column density and temperature maps.
Fifteen {\it Herschel} clumps (having M$_{clump}$ $\sim$15 to 1875 M$_{\odot}$) are traced in the filamentary cloud. 
Most of the parts of the cloud is depicted in a temperature range of $\sim$10--12~K, where the presence of the residual amounts of very cold H\,{\sc i} gas is also evident. However, the most massive clump (having peak N(H$_{2}$) $\sim$4.8 $\times$ 10$^{22}$ cm$^{-2}$; A$_{V}$ $\sim$51 mag) is traced in a temperature range of 18--26 K.
Several noticeable parsec-scale {\it Herschel} filaments appear to be radially pointed to this most massive clump. 
It can be referred to as a ``hub-filament" configuration. The intense star formation activities are also observed toward the most massive {\it Herschel} clump (see Section~\ref{subsec:srf} and also the surface density contours in Figure~\ref{fs9}b), which is also not associated with any ionized emission. 
A majority of YSOs and their clustering are found toward this massive clump (or a rich cluster clump), hence, a relatively high value of temperature 
toward the clump can be explained by the presence of star formation activities. 
One can also note that the clump also harbors an IRc of the 6.7 GHz MME that could be 
a MYSO candidate prior to the UCH\,{\sc ii} phase (see Section~\ref{sec:clmtv}). 

All together, the cluster of YSOs (including a MYSO candidate) is located at the ``hub-filament" (i.e. the most massive clump) system. 
Interestingly, a similar morphology, where the infrared clusters are located at the junction of filaments or filament mergers, has been found in other cloud complexes, such as Taurus, Ophiuchus, Rosette, SDC335.579-0.292, W42, and Sh 2-138 etc. \citep[e.g.][]{myers09,schneider12,peretto13,dewangan15,baug15}. In the simulations of \citet{dale11}, 
a similar result was reported. \citet{myers09} also presented a theoretical scenario based on nearby star-forming 
complexes of star clusters forming within ``hubs" and suggested that the ``hub region" can be associated 
with a peak N(H$_{2}$) value of $\sim$3 $\times$ 10$^{22}$ cm$^{-2}$, which is also observed in the I05480+2545 site. 
It is likely that the filaments might funnel material to the central hub system and the clump/core, which also contains a MYSO at its initial stage 
(including young stellar cluster). 
To confirm this indicative argument, a thorough investigation of the I05480+2545 site is encouraged using high-resolution molecular line observations.
\section{Summary and Conclusions}
\label{sec:conc}
In this paper, we have examined the physical environment and the distribution of YSOs in our selected region around I05480+2545, 
using the multi-wavelength data spanning from optical H$\alpha$, NIR to radio wavelengths. 
The main outcomes of our study are the following:\\
$\bullet$ The I05480+2545 site is embedded in an elongated filamentary cloud evident in the {\it Herschel} column density (N(H$_{2}$)) map. 
Using the $^{13}$CO intensity map, a similar morphology was also reported earlier by \citet{kawamura98}. 
Fifteen {\it Herschel} clumps (having M$_{clump}$ $\sim$15 to 1875 M$_{\odot}$) are identified in the elongated filamentary cloud. 
The filamentary cloud also harbors the H\,{\sc ii} region IRAS 05480+2544/BFS 48, where an inverted V-like morphology is also 
seen in the multi-wavelength images.\\  
$\bullet$ The most massive clump (having peak N(H$_{2}$) $\sim$4.8 $\times$ 10$^{22}$ cm$^{-2}$; A$_{V}$ $\sim$51 mag) contains the 6.7 GHz MME 
and is traced in a temperature range of 18--26 K. However, the remaining filamentary cloud is depicted in a temperature range of $\sim$10--12~K, 
where the HISA features are also found, indicating the presence of the residual amounts of very cold H\,{\sc i} gas.\\ 
$\bullet$ An IRc of the 6.7 GHz MME is found in the {\it Spitzer} 3.6--4.5 $\mu$m images (resolution $\sim$2$\arcsec$). 
The IRc does not look like a point-like source and seems to be linked with the molecular outflow. 
Based on the H$\alpha$ and NVSS 1.4 GHz images, the ionized emission is absent toward the IRc. 
Together, the most massive clump harbors an early phase of MYSO candidate before the onset of the UCH\,{\sc ii} region.\\ 
$\bullet$ Several noticeable parsec-scale filaments are evident in the {\it Herschel} image at 250 $\mu$m and 
seem to be radially pointed to the most massive clump. It represents a ``hub-filament" system. \\ 
$\bullet$ 181 YSOs are identified in the selected region through the analysis of NIR photometric data at 1--5 $\mu$m. 
The intense star formation activities are investigated toward the most massive {\it Herschel} clump.

Taken together, the cluster of YSOs (including a MYSO candidate) is found at the junction (i.e. the most massive clump) of the filaments.
It appears a very similar system to those seen in Rosette Molecular Cloud. 
These results favor the role of filaments in the formation of the massive star-forming clump and young stellar cluster in the I05480+2545 site. 
We also conclude that the I05480+2545 site contains the early phase of massive star formation 
before the onset of a hyper-compact(HC)/UCH\,{\sc ii} phase. 
\acknowledgments 
We thank the anonymous reviewer for constructive comments and suggestions. 
The research work at Physical Research Laboratory is funded by the Department of Space, Government of India.
We acknowledge the local staff at the Mount Abu InfraRed Observatory (MIRO) for their help. 
This work is based on data obtained as part of the UKIRT Infrared Deep Sky Survey. This publication 
made use of data products from the Two Micron All Sky Survey (a joint project of the University of Massachusetts and 
the Infrared Processing and Analysis Center / California Institute of Technology, funded by NASA and NSF), archival 
data obtained with the {\it Spitzer} Space Telescope (operated by the Jet Propulsion Laboratory, California Institute 
of Technology under a contract with NASA). The Dominion Radio Astrophysical Observatory is operated as a national facility by the 
National Research Council of Canada. The Canadian Galactic Plane Survey (CGPS) has been a Canadian project with 
international partners, and was supported by grants from the Natural Sciences and Engineering Research Council of Canada (NSERC). 
This paper makes use of data obtained as part of the INT Photometric H$\alpha$ Survey of the Northern Galactic 
Plane (IPHAS, www.iphas.org) carried out at the Isaac Newton Telescope (INT). The INT is operated on the 
island of La Palma by the Isaac Newton Group in the Spanish Observatorio del Roque de los Muchachos of 
the Instituto de Astrofisica de Canarias. The IPHAS data are processed by the Cambridge Astronomical Survey 
Unit, at the Institute of Astronomy in Cambridge. 
%
%\clearpage
%
%
%%\clearpage

%\newpage
%
\begin{deluxetable}{ccccc}
\tablewidth{0pt} 
\tabletypesize{\scriptsize} 
\tablecaption{Physical properties of the {\it Herschel} clumps detected in our selected field around I05480+2545 (see Figures~\ref{fs6}b and~\ref{fs6}c). 
Column~1 gives the IDs assigned to the clump. Table also lists 
positions, deconvolved effective radius (R$_{c}$), and clump mass (M$_{clump}$). The clump (ID \#4) highlighted with a dagger contains a cluster of YSOs 
and the 6.7 GHz MME. \label{tab1}} 
\tablehead{ \colhead{ID} & \colhead{{\it l}} & \colhead{{\it b}} & \colhead{R$_{c}$}& \colhead{M$_{clump}$}\\
\colhead{} &  \colhead{[degree]} & \colhead{[degree]} & \colhead{(pc)} &\colhead{($M_\odot$)}}
\startdata 
   1   &     183.417	&   -0.607	&    1.6  &  	 545	\\	
   2   &     183.390	&   -0.502	&    1.0  &  	 135	\\	
   3   &     183.351	&   -0.537	&    1.1  &  	 240	\\		  
   4$\dagger$   &     183.355	&   -0.580	&    1.7  &  	1875	\\		
   5   &     183.327	&   -0.514	&    0.8  &  	 105	\\		 
   6   &     183.273	&   -0.584	&    1.6  &  	 595	\\		
   7    &    183.242	 &  -0.623	 &   0.8   & 	  90	\\		 
   8   &     183.215	&   -0.588	&    0.9  &  	 125	\\		
   9   &     183.168	&   -0.603	&    0.9  &  	 130	\\		
  10   &     183.117	&   -0.689	&    1.5  &  	 335	\\		 
  11   &     183.141	&   -0.786	&    1.6  &  	 385	\\		
  12    &    183.203	 &  -0.661	 &   0.6   & 	  40	\\	 
  13   &     183.230	&   -0.755	&    0.4  &  	  15	\\		
  14   &     183.110	&   -0.599	&    0.5  &  	  40	\\		  
  15    &    183.028	 &  -0.693	 &   0.5   & 	  35	\\		
 \enddata  
\end{deluxetable}


\begin{thebibliography}{}
%
\bibitem[Aguirre et al.(2011)]{aguirre11}
Aguirre, J.~E., Ginsburg, A.~G., Dunham, M.~K., et al. 2011, ApJS, 192, 4

\bibitem[Assirati et al.(2014)]{assirati14}
Assirati, L., Silva, N.~R., Berton, L., Lopes, A.~A., \& Bruno, O.~M. 2014, Journal of Physics: Conference Series, 490(1), 2014

\bibitem[Anandarao et al.(2008)]{anandarao08}
Anandarao, B.~G., Richardson, E.~H., Chakraborty, A. \& Epps, H. 2008, 
Ground-based and Airborne Instrumentation for Astronomy II, 
Edited by McLean, Ian S.; Casali, Mark M. Proceedings of the SPIE, Volume 7014, pp. 70142Y-70142Y-8

\bibitem[Andr{\'e} et al.(2010)]{andre10}
Andr{\'e}, P., Men'shchikov, A., Bontemps, S., et al. 2010, A\&A, 518, L102

\bibitem[Andr{\'e} et al.(2016)]{andre16}
Andr{\'e}, P., Rev{\'e}ret, V., K\"{o}nyves, V., et al. 2016, A\&A, 592, 54

\bibitem[Baug et al.(2015)]{baug15}
Baug, T., Ojha, D.~K., Dewangan, L.~K., et al. 2015, MNRAS, 454, 4335

\bibitem[Blitz et al.(1982)]{blitz82}
Blitz, L., Fich, M., \& Stark, A.~A. 1982,  ApJS, 49, 183

\bibitem[Bohlin et al.(1978)]{bohlin78}
Bohlin, R.~C., Savage, B.~D., \& Drake, J.~F. 1978, ApJ, 224, 13233

\bibitem[Bressert et al.(2010)]{bressert10}
Bressert, E., Bastian, N., Gutermuth, R., et al. 2010, MNRAS, 409, 54

\bibitem[Condon et al.(1998)]{condon98}
Condon, J.~J., Cotton, W.~D., Greisen, E.~W., et al. 1998, AJ, 115, 1693

\bibitem[Contreras et al.(2016)]{contreras16}
Contreras, Y., Garay, G., Rathborne, J.~M., \& Sanhueza,P. 2016, MNRAS, 456, 2041

\bibitem[Dale \& Bonnell(2011)]{dale11}
Dale, J.~E.,  \& Bonnell, I.~A. 2011, MNRAS, 414, 321

\bibitem[Dewangan et al.(2015a)]{dewangan15a}
Dewangan, L.~K., Mayya, Y.~D., Luna, A., \& Ojha, D.~K.  2015a, ApJ, 803, 100

\bibitem[Dewangan et al.(2015b)]{dewangan15}
Dewangan, L.~K., Luna, A., Ojha, D.~K., et al.  2015b, ApJ, 811, 79

\bibitem[Dewangan et al.(2016a)]{dewangan16}
Dewangan, L.~K., Ojha, D.~K., Luna, A., et al.  2016a, ApJ, 819, 66

\bibitem[Dewangan et al.(2016b)]{dewangan16b}
Dewangan, L.~K., Ojha, D.~K., Zinchenko, I., et al.  2016b, ApJ, 833, 246

\bibitem[Dewangan et al.(2017)]{dewangan17a}
Dewangan, L.~K., Ojha, D.~K., Zinchenko, I., Janardhan, P., \& Luna, A.  2017, ApJ, 834, 22

\bibitem[Drew et al.(2005)]{drew05}
Drew, J.~E., Greimel, R., Irwin, M.J., et al. 2005, MNRAS, 362, 753

\bibitem[Flaherty et al.(2007)]{flaherty07}
Flaherty, K.~M., Pipher, J.~L., Megeath, S.~T., et al. 2007, ApJ, 663, 1069

\bibitem[Gonzalez \& Woods(2011)]{gonzalez11}
Gonzalez, R, \& Woods, R. 2011, {\it Digital Image Processing} (Pearson Education) ISBN 9780133002324

\bibitem[Griffin et al.(2010)]{griffin10} 
Griffin, M.~J., Abergel, A., Abreu, A, et al. 2010, A\&A, 518L, 3

\bibitem[Gutermuth et al.(2009)]{gutermuth09}
Gutermuth, R.~A., Megeath, S.~T., Myers, P.~C., et al. 2009, ApJS, 184, 18

\bibitem[Henning et al.(1992)]{henning92}
Henning, Th., Cesaroni, R., Walmsley, M., \& Pfau, W. 1992, A\&AS, 93, 525

\bibitem[Hildebrand(1983)]{hildebrand83} 
Hildebrand, R.~H. 1983, QJRAS, 24, 267

\bibitem[Kainulainen et al.(2016)]{kainulainen16}
Kainulainen, J., Hacar, A.,  Alves, J., et al. 2016, A\&A, 586,27

\bibitem[Kauffmann et al.(2008)]{kauffmann08}
Kauffmann, J., Bertoldi, F., Bourke, T.~L., Evans, II, N.~J.,\&  Lee, C.~W. 2008, ApJ, 487, 993

\bibitem[Kawamura et al.(1998)]{kawamura98}
Kawamura, A., Onishi, T., Yonekura, Y., et al. 1998, ApJS, 117, 387

\bibitem[Kerton(2005)]{kerton05}
Kerton, C.~R. 2005, ApJ, 623, 235

\bibitem[Klein et al.(2005)]{klein05}
Klein, R.,  Posselt, B., Schreyer, K., Forbrich, J., \& Henning, Th. 2005, ApJS, 161, 361

\bibitem[Lawrence et al.(2007)]{lawrence07}
Lawrence, A., Warren, S.~J., Almaini, O., et al. 2007, MNRAS, 379, 1599

\bibitem[Li et al.(2016)]{li16}
Li, Guang-Xing, Urquhart, J.~S., Leurini, S., et al. 2016, A\&A, 591, 5

\bibitem[Mallick et al.(2015)]{mallick15}
Mallick, K.~K., Ojha, D.~K., Tamura, M., et al. 2015, MNRAS, 447, 2307

\bibitem[MacLeod et al.(1998)]{macleod98}
MacLeod, G.~C., Scalise, Jr., E., Saedt, S., Galt, J. A., \& Gaylard, M. J. 1998, AJ, 116, 1897

\bibitem[Minier et al.(2001)]{minier01}
Minier, V., Conway, J.~E., \& Booth, R.~S. 2001, A\&A, 369, 278

\bibitem[Myers (2009)]{myers09} 
Myers, P.~C. 2009, ApJ, 700, 1609

\bibitem[Nakamura et al.(2014)]{nakamura14}
Nakamura, F., Sugitani, K., Tanaka, T., et al. 2014, ApJL, 791, L23

\bibitem[Ott(2010)]{ott10}
Ott, S. 2010, in Astronomical Society of the Pacic Conference
Series, Vol. 434, Astronomical Data Analysis Software and
Systems XIX, ed. Y. Mizumoto, K.-I. Morita, \& M. Ohishi, 139

\bibitem[Peretto et al.(2013)]{peretto13}	
Peretto, N., Fuller, G.`A., Duarte-Cabral, A., et al. 2013, A\&A, 555, 112

\bibitem[Poglitsch et al.(2010)]{poglitsch10}	
Poglitsch, A., Waelkens, C., Geis, N., et al. 2010, A\&A, 518L, 2

\bibitem[Ragan et al.(2014)]{ragan14}
Ragan,S.~E., Henning, Th., Tackenberg, J., et al. 2014, A\&A, 568, 73

\bibitem[Schneider et al.(2012)]{schneider12}
Schneider, N., Csengeri, T., Hennemann, M., et al. 2012, A\&A, 540, L11

\bibitem[Skrutskie et al.(2006)]{skrutskie06}
Skrutskie, M.~F., Cutri, R.~M., Stiening, R., et al. 2006, AJ, 131, 1163

\bibitem[Slysh et al.(1997)]{slysh97}
Slysh, V.~I., Dzura, A.~M., Val'tts, I.~E., Gerard, E. 1997, A\&AS,124, 85

\bibitem[Sunada et al.(2007)]{sunada07} 	
Sunada, K., Nakazato, T., Ikeda, N., et al. 2007, PASJ, 59, 1185

\bibitem[Szymczak et al.(2012)]{szymczak12} 	
Szymczak, M., Wolak, P., Bartkiewicz, A., \& Borkowski, K.~M. 2012, AN, 333, 634

\bibitem[Tan et al.(2014)]{tan14} 	
Tan, J.~C., Beltr\'an, M.~T., Caselli, P., et al. 2014, in Protostars and Planets VI, ed. H. Beuther et al. (Tucson, AZ: Univ. Arizona Press), 149

\bibitem[Taylor et al.(2003)]{taylor03}
Taylor, A.~R., Gibson, S.~J., Peracaula, M., et al. 2003, AJ, 125, 3145

\bibitem[Urquhart et al.(2013)]{urquhart13} 
Urquhart, J.~S., Moore, T.~J.~T., Schuller, F., et al. 2013, MNRAS, 431, 1752

\bibitem[Walsh et al.(1998)]{walsh98}
Walsh, A.~J., Burton, M.~G., Hyland, A.~R., \& Robinson, G. 1998, MNRAS, 301, 640

\bibitem[Whitney et al.(2011)]{whitney11}
Whitney, B., Benjamin, R., Meade, M., et al. 2011, BAAS, 43, 241.16

\bibitem[Williams et al.(1994)]{williams94} 
Williams, J. P., de Geus, E. J., \& Blitz, L. 1994, ApJ, 428, 693

\bibitem[Wright et al.(2010)]{wright10}
Wright, E.~L., Eisenhardt, P.~R.~M., Mainzer, A.~K., et al. 2010, AJ, 140, 1868

\bibitem[Wu et al.(2010)]{wu10}
Wu, Y.~W., Xu, Y., Pandian, J.~D., et al. 2010, ApJ, 720, 392

\bibitem[Wu et al.(2011)]{wu11}
Wu, Y.~W., Xu, Y., Yang, J. 2011, Research in Astronomy and Astrophysics, 11, 137
%
\end{thebibliography}
 \end{document}